\documentclass[twocolumn]{aastex701}

\usepackage{apjfonts}

\newcommand{\lum}{erg\,s\ensuremath{^{-1}}}
\usepackage{multirow}

\begin{document}

\title{EP~J062132.0+623837: A Multi-wavelength Study of a Changing-look Active Galactic Nucleus Discovered by Einstein Probe}

\author[0009-0006-4524-4414]{Lin-Na Xu}
\affiliation{Department of Astronomy, University of Science and Technology of China, Hefei, 230026, China}
\affiliation{School of Astronomy and Space Sciences,
University of Science and Technology of China, Hefei, 230026, China}
\email[show]{xulinna@mail.ustc.edu.cn} 

\author[0000-0002-7152-3621]{Ning Jiang}
\affiliation{Department of Astronomy, University of Science and Technology of China, Hefei, 230026, China}
\affiliation{School of Astronomy and Space Sciences,
University of Science and Technology of China, Hefei, 230026, China}
\email[show]{jnac@ustc.edu.cn} 

\author[0000-0002-1517-6792]{Tinggui Wang}
\affiliation{Department of Astronomy, University of Science and Technology of China, Hefei, 230026, China}
\affiliation{School of Astronomy and Space Sciences, University of Science and Technology of China, Hefei, 230026, China}
\affiliation{Department of Physics and Astronomy, College of Physics, Guizhou University, Guiyang 550025, People's Republic of China}
\email[]{twang@ustc.edu.cn} 

\author[0000-0001-7689-6382]{Shifeng Huang}
\affiliation{Department of Astronomy, University of Science and Technology of China, Hefei, 230026, China}
\affiliation{School of Astronomy and Space Sciences, University of Science and Technology of China, Hefei, 230026, China}
\email[]{sfhuang999@ustc.edu.cn} 

\author[0000-0003-4225-5442]{Yibo~Wang}
\affiliation{Department of Astronomy, University of Science and Technology of China, Hefei, 230026, China}
\affiliation{School of Astronomy and Space Sciences, University of Science and Technology of China, Hefei, 230026, China}
\email[]{wybustc@ustc.edu.cn} 

\author[0000-0001-6938-8670]{Zhenfeng Sheng}
\affiliation{Institute of Deep Space Sciences, Deep Space Exploration Laboratory, Hefei, 230026, People’s Republic of China}
\email[]{shengzf@ustc.edu.cn} 

\author[0009-0006-8112-0187]{Mingxin Wu}
\affiliation{Department of Astronomy, University of Science and Technology of China, Hefei, 230026, China}
\affiliation{School of Astronomy and Space Sciences,
University of Science and Technology of China, Hefei, 230026, China}
\email[]{wmx112611226@mail.ustc.edu.cn} 

\author[0000-0003-3824-9496]{Jiazheng Zhu}
\affiliation{Department of Astronomy, University of Science and Technology of China, Hefei, 230026, China}
\affiliation{School of Astronomy and Space Sciences, University of Science and Technology of China, Hefei, 230026, China}
\email[]{jiazheng@mail.ustc.edu.cn} 

\author[0000-0003-4200-9954]{Huaqing Cheng}
\affiliation{National Astronomical Observatories, Chinese Academy of Sciences, Beijing, 100101, China}
\email[]{hqcheng@nao.cas.cn} 

\author[0000-0002-2006-1615]{Chichuan Jin}
\affiliation{National Astronomical Observatories, Chinese Academy of Sciences, Beijing, 100101, China}
\email[]{ccjin@nao.cas.cn} 

\author[0000-0002-8402-3722]{Jun-Jie Jin}
\affiliation{National Astronomical Observatories, Chinese Academy of Sciences, Beijing, 100101, China}
\email[]{jjjin@bao.ac.cn} 

\author[0000-0002-4562-7179]{Dongyue Li}
\affiliation{National Astronomical Observatories, Chinese Academy of Sciences, Beijing, 100101, China}
\email[]{dyli@nao.cas.cn}

\author[0009-0007-6104-337X]{Yuan Liu}
\affiliation{National Astronomical Observatories, Chinese Academy of Sciences, Beijing, 100101, China}
\email[]{liuyuan@bao.ac.cn} 

\author[0000-0002-6221-1829]{Jianwei Lyu}
\affiliation{Steward Observatory, University of Arizona, 933 North Cherry Avenue, Tucson, AZ 85721, USA}
\email[]{jianwei@arizona.edu} 

\author[0000-0003-4156-3793]{Jian-Guo Wang}
\affiliation{Yunnan Observatories, Chinese Academy of Sciences, Kunming 650011, PR China}
\affiliation{Key Laboratory for the Structure and Evolution of Celestial Objects, Yunnan Observatories, Kunming 650011, China}
\email[]{wangjg@ynao.ac.cn} 

\author[]{Yun-Fei Xu}
\affiliation{National Astronomical Observatories, Chinese Academy of Sciences, Beijing, 100101, China}
\email[]{xuyf@nao.cas.cn} 

\author[0000-0002-1935-8104]{Yongquan Xue}
\affiliation{Department of Astronomy, University of Science and Technology of China, Hefei, 230026, China}
\affiliation{School of Astronomy and Space Sciences,
University of Science and Technology of China, Hefei, 230026, China}
\email[]{xuey@ustc.edu.cn} 

\author[0000-0002-3403-8002]{Zhen Zhang}
\affiliation{National Astronomical Observatories, Chinese Academy of Sciences, Beijing, 100101, China}
\email[]{zhangzhen@nao.cas.cn}


\begin{abstract}

EP~J062132.0+623837 is a new X-ray source identified in 14-day binned survey data from the Wide-field X-ray Telescope aboard the Einstein Probe (EP). Subsequent X-ray observations with the Follow-up X-ray Telescope (FXT) on board EP localized the source to the central region of the galaxy 2MASX~J06213207+6238390 at $z=0.084$. Here, we present a comprehensive analysis of the multi-wavelength data for this object. During the FXT and \textit{Swift}/XRT follow-up, the source remained X-ray active, with observed luminosities of $(3-9)\times10^{43}$~\lum, and its X-ray spectra were well described by power-law models with photon indices of $\Gamma\sim1.5$. The optical spectrum of the host galaxy clearly reveals a broad-line active galactic nucleus (AGN) exhibiting significantly enhanced broad Balmer emission lines compared to an archival spectrum obtained four years earlier, thereby marking a changing-look transition from Type~1.8 to Type~1.5. Furthermore, long-term optical and mid-infrared light curves indicate that the nucleus underwent dramatic brightening around 2010 and has since shown substantial variability. Given the spatial coincidence and multi-wavelength behavior, we identify EP~J062132.0+623837 as a changing-look AGN, initially detected by EP and subsequently confirmed to exhibit large-amplitude variability across multiple bands. This case study demonstrates the great potential of wide-field time-domain X-ray surveys for discovering changing-look AGNs, paving a new avenue for their systematic identification as a powerful complement to ongoing optical surveys.
\end{abstract}


\section{Introduction}

Changing-look active galactic nuclei (CLAGNs) are a remarkable class of accreting supermassive black holes that undergo dramatic spectral-state changes on humanly accessible timescales. Their defining observational signature is the appearance, disappearance, or strong change of the broad emission-line component, often accompanied by substantial continuum variability across multiple wavebands \citep[e.g.,][]{LaMassa2015,MacLeod2016}. CLAGNs provide a valuable probe of the inner accretion flow, the ionizing radiation field, and the response of the broad-line region (BLR) and circumnuclear dust, and have therefore become an important topic in time-domain studies of black hole accretion \citep{RicciTrakhtenbrot2023}.  Large optical time-domain surveys have further shown that CLAGNs are not isolated curiosities, but part of a broader population of strongly variable AGNs uncovered in recent years \citep[e.g.,][]{Green2022,Yang2023}.

The changing-look phenomenon encompasses a wide range of spectroscopic behaviors. In the most dramatic cases, an AGN transitions from a Type~1 spectrum, characterized by prominent broad Balmer emission lines, to a Type~2 spectrum dominated by narrow emission lines \citep[e.g.,][]{LaMassa2015,Ruan2016}, or vice versa \citep[e.g.,][]{Gezari2017ApJ,Yan2019}. In low-redshift AGNs, changing-look behavior can also include transitions among intermediate Seyfert subclasses, such as Types 1.2, 1.5, 1.8, and 1.9. These classifications are traditionally based on the visibility and relative strength of the broad Balmer components \citep[e.g.,][]{Osterbrock1977,Osterbrock1981,OsterbrockFerland2006}, and can in some cases be quantified using empirical flux ratios involving H$\beta$, H$\alpha$, and [O\,\textsc{iii}] \citep[e.g.,][]{Cohen1986,Winkler1992}. Recent studies have increasingly recognized such intermediate-type transitions as part of the broader changing-look AGN (CLAGN) phenomenon \citep[e.g.,][]{Denney2014,Li2022,Guo2024DESI,RicciTrakhtenbrot2023}. In this sense, CLAGNs are best understood as manifestations of dramatic AGN state changes, rather than as a narrowly defined class of events \citep{RicciTrakhtenbrot2023}.

Most CLAGNs have so far been discovered through optical routes. One major path has been the identification of large-amplitude optical variability in wide-field surveys, followed by spectroscopic confirmation of the changing-look nature of the source \citep[e.g.,][]{MacLeod2016,Green2022}. Another has been the direct comparison of repeated optical spectra, which remains one of the most straightforward ways to establish changes in the broad-line appearance \citep[e.g.,][]{LaMassa2015,Yang2018}. Systematic multi-epoch spectroscopic searches have further expanded the CLAGN census, including recent DESI-based work that has substantially increased the known sample and explicitly incorporated intermediate-type transitions into the broader CLAGN search framework \citep{Guo2024DESI}. These optical approaches have produced the majority of currently known CLAGN samples and have established the phenomenological diversity of the class \citep[e.g.,][]{Yang2023,RicciTrakhtenbrot2023}.

Mid-infrared (MIR) variability provides a complementary and increasingly important discovery route. Because the hot-dust emission traces reprocessed radiation from the central engine and is less sensitive to dust extinction than the optical continuum, MIR light curves can help distinguish intrinsic accretion-state changes from simple obscuration scenarios \citep[e.g.,][]{Sheng2017,Lyu2022}. In addition, systematic searches based on large-amplitude MIR variability and MIR color evolution have proved effective in selecting CLAGN candidates, opening a selection channel that is especially valuable for host-dominated or dust-affected systems \citep{Sheng2020,Jiang2021}.

In principle, the X-ray band should be one of the most sensitive tracers of changes in black-hole accretion activity, since it directly probes the innermost accretion flow and corona \citep{RicciTrakhtenbrot2023}. Wide-area X-ray surveys such as eROSITA have dramatically improved the census of X-ray sources and variable AGNs \citep{Merloni2024}. Indeed, eROSITA has already been used to discover a new changing-look Seyfert galaxy \citep{Markowitz2026}. However, the systematic discovery of CLAGNs from X-ray data has remained comparatively limited, because existing survey strategies are generally better suited to all-sky coverage and long-term variability studies than to the frequent, low-latency identification and immediate follow-up of changing-look behavior across the sky. In this respect, Einstein Probe (EP, \citealt{Yuan2022,Yuan2025}), with repeated soft X-ray monitoring by the Wide-field X-ray Telescope (WXT) and rapid follow-up by the Follow-up X-ray Telescope (FXT), opens a new opportunity for identifying changing-look AGNs from the X-ray band.

In this paper, we present a multiwavelength study of EP~J062132.0+623837, an X-ray source associated with the galaxy 2MASX~J06213207+6238390 at \(z = 0.084011 \pm 0.000001\), as measured by DESI DR1 \citep{DESIDR1}, first identified by WXT in 14-day binned data. Follow-up observations with FXT localized the source to the nucleus of its host galaxy, while subsequent X-ray monitoring confirmed its persistent activity. Long-term optical and mid-infrared light curves reveal pronounced multiwavelength variability extending several years prior to the EP trigger. Moreover, the strengthening of broad Balmer emission lines between pre- and post-discovery optical spectra indicates a transition from a Type~1.8 to a Type~1.5 AGN, identifying EP~J062132.0+623837 as a changing-look AGN.


This paper is organized as follows. In Section~\ref{sec:2}, we describe the X-ray, ultraviolet, optical, IR, and spectroscopic data sets and summarize the data reduction procedures. In Section~\ref{sec:3}, we present the multiwavelength light curves, the X-ray properties, and the optical spectral results. In Section~\ref{sec:4}, we discuss the physical interpretation of the source, its relation to changing-look AGNs, and the implications for EP nuclear transients. Our main conclusions are summarized in Section~\ref{sec:5}. In this work, we assume cosmological parameters of $H_0=70\,\text{km}\,\text{s}^{-1}\,\text{Mpc}^{-1}$, $\Omega_{\rm M}=0.3$, and $\Omega_{\Lambda}=0.7$.

\section{Observations and Data Reduction}\label{sec:2}

\subsection{WXT Observations}

EP~J062132.0+623837 was identified as a variable X-ray source in survey data of EP-WXT, the wide-field telescope employing novel lobster-eye micro-pore focusing technology in the soft X-ray band. We retrieved the standard WXT light-curve products by querying the WXT photon database with a positional matching radius of $3'$ centered on EP~J062132.0+623837, and constructed 14-day binned measurements over the interval from 2025 January 1 to 2026 April 4. The standard WXT light-curve products provide source and background light curves extracted from predefined regions. The source light curve is extracted using a circular aperture with a radius of approximately \(9.1^\prime\), while the background is estimated from a concentric annulus with inner and outer radii of approximately \(18.2^\prime\) and \(36.4^\prime\), respectively. The background region has an area approximately 12 times that of the source region, and the scaled background light curve was subtracted from the source light curve using the corresponding area scaling factor of \(1/12\). The resulting net light curve was then rebinned into 14-day intervals. The source is weak in individual WXT snapshots and in many of the 14-day bins, so the temporal binning was adopted to improve the sensitivity to low-level long-timescale variability. The accumulated WXT exposure varies substantially among the 14-day bins as a consequence of the survey observing pattern, and this variation affects the statistical precision and detection significance of individual bins.

The WXT light curve was evaluated in the 0.5--4.0 keV band. To obtain approximate flux estimates, we converted the 14-day binned count rates to fluxes using an empirical conversion factor of \(2\times10^{-9}\ \mathrm{erg\ cm^{-2}\ count^{-1}}\), adopted here only for order-of-magnitude characterization of the WXT detections in the context of the current instrumental calibration \citep[e.g.,][]{Cheng2025,Cheng2026}. In the absence of significant detections, the resulting measurements are treated as indicative flux constraints rather than as precise flux determinations.

\begin{figure}
\centering
\includegraphics[width=0.45\textwidth]{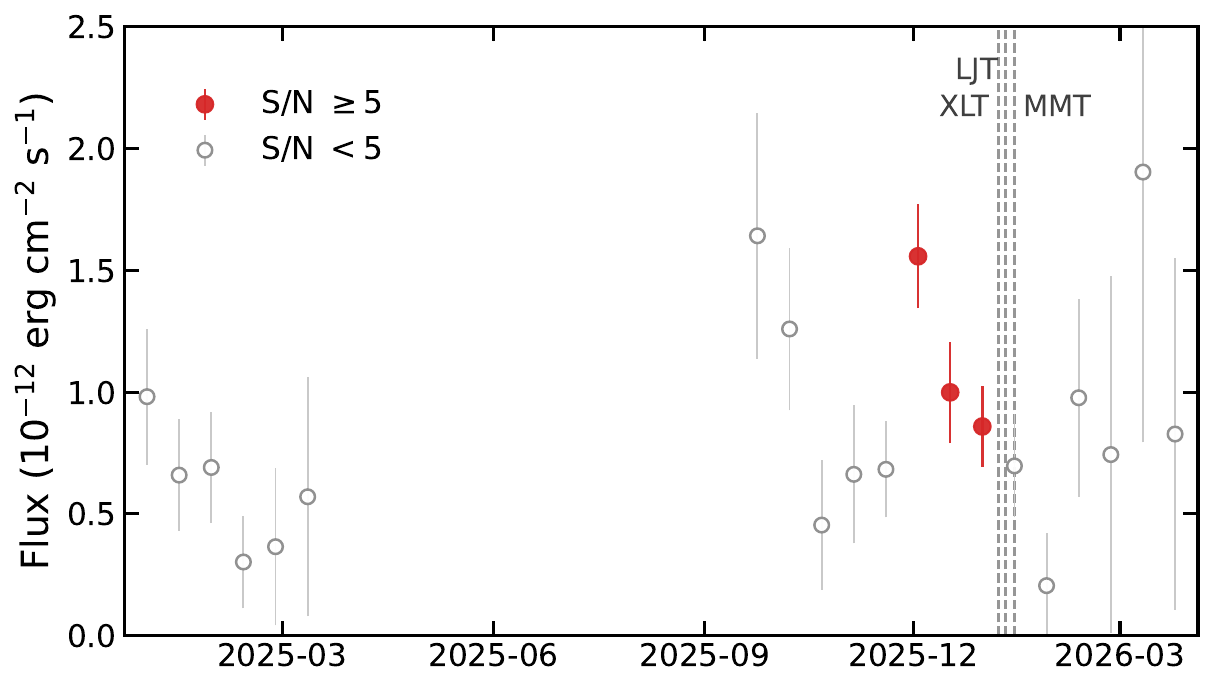}
\caption{Fourteen-day binned WXT light curve of EP~J062132.0+623837 in the 0.5--4.0 keV band, shown in approximate observed flux derived using the empirical count-rate-to-flux conversion described in the text. Filled and open symbols denote bins with detection significance above and below \(5\sigma\), respectively. The vertical dashed lines mark the XLT, LJT, and MMT spectroscopic follow-up observations on 2026 January 7, 2026 January 10, and 2026 January 14. The archival DESI spectrum was obtained on 2021 December 15.
}
\label{fig:wxt_discovery}
\end{figure}

Figure~\ref{fig:wxt_discovery} shows the 14-day binned WXT light curve of EP~J062132.0+623837 in the 0.5--4.0 keV band, displayed in approximate observed flux using the empirical count-rate-to-flux conversion described above. The source is weak in most time bins, but becomes clearly detectable in 2025 December. Three adjacent 14-day bins exceed \(5\sigma\), and the highest-significance bin reaches \(8.46\sigma\). The strongest bin corresponds to an approximate observed flux of \((1.56\pm0.21)\times10^{-12}\ \mathrm{erg\ cm^{-2}\ s^{-1}}\). Earlier bins generally have lower detection significance, although the accumulated exposure varies substantially with epoch and therefore affects the sensitivity of individual bins. We therefore do not interpret changes in WXT detection significance alone as evidence for intrinsic long-term flux evolution.

\subsection{FXT Follow-up Observation}

Following the WXT discovery, EP performed a follow-up observation with the FXT on 2026 January 5 (MJD 61045.886). The observation was obtained with the two FXT units, each with an effective exposure of 4.27 ks. A source was clearly detected in the FXT field near the nucleus of the host galaxy 2MASX J06213207+6238390. We processed the data using the Follow-up X-ray Telescope Data Analysis Software (\texttt{FXTDAS}) v1.30, employing the task \texttt{fxtchain} to generate light curves and spectra. Source region was extracted from a circular region of radius $60^{\prime\prime}$ centered on the target, while the background region was taken from a nearby, source-free circular region with a radius of $200^{\prime\prime}$.

\begin{figure}
\centering
\includegraphics[width=0.45\textwidth]{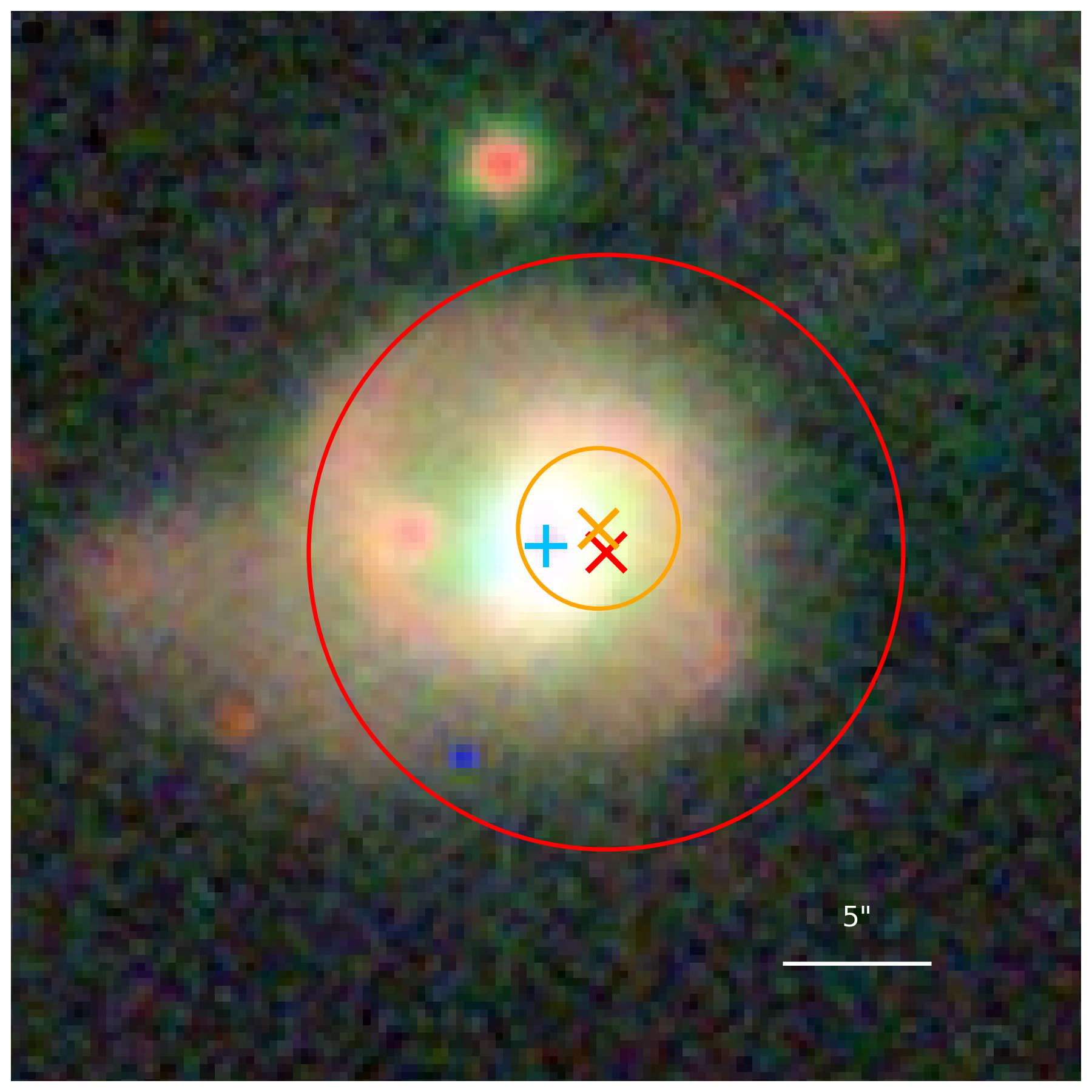}
\caption{
X-ray localizations of EP~J062132.0+623837 overlaid on the DESI Legacy Imaging Surveys color image of the host galaxy. The blue cross marks the host-galaxy nucleus. The red cross and circle show the refined EP/FXT centroid and its 90\% uncertainty region with a radius of \(10\arcsec\), while the orange cross and circle show the UKSSDC enhanced \textit{Swift}/XRT position and its 90\% uncertainty region with a radius of \(2.7\arcsec\). The FXT and enhanced XRT positions are offset from the host nucleus by \(2.03^{\prime\prime}\) and \(1.85^{\prime\prime}\), respectively. The scale bar corresponds to 5\arcsec.
}
\label{fig:fxt_loc}
\end{figure}

The refined FXT centroid is measured at \(\alpha = 06^{\mathrm h}21^{\mathrm m}31.872^{\mathrm s}\), \(\delta = +62^\circ38^\prime38.4^{\prime\prime}\) (J2000), with a 90\% positional uncertainty radius of \(10^{\prime\prime}\). The host-galaxy nuclear position measured from the Dark Energy Spectroscopic Instrument (DESI) Legacy Imaging Surveys image is located at \(\alpha = 06^{\mathrm h}21^{\mathrm m}32.165^{\mathrm s}\), \(\delta = +62^\circ38^\prime38.608^{\prime\prime}\) (J2000). The angular separation between the FXT centroid and the host-galaxy nucleus is \(2.03^{\prime\prime}\). Figure~\ref{fig:fxt_loc} shows the FXT localization on the DESI color image \citep{Dey2019}. The enhanced \textit{Swift}/XRT position derived below is also shown for comparison.

\subsection{\textit{Swift} X-Ray Observations}

We also obtained follow-up observations with the \textit{Neil Gehrels Swift Observatory} \citep{Gehrels2004}. We observed the source with the X-Ray Telescope \citep[XRT;][]{Burrows2005} at four epochs between 2026 January 11 and February 8. The data were processed using the \texttt{HEASoft v6.36} tasks \texttt{xrtpipeline} and \texttt{xrtproducts} to generate X-ray light curves and spectra. Source counts were extracted from a circular region with a radius of $40^{\prime\prime}$ centered on the target, while the background was from an annular region with inner and outer radii of $60^{\prime\prime}$ and $180^{\prime\prime}$, respectively. The UK Swift Science Data Centre XRT products builder was used to derive an
enhanced \textit{Swift}/XRT position, with the XRT astrometry improved using
UVOT field astrometry \citep[e.g.,][]{Goad2007,Evans2009}. The resulting position is
$\alpha=06^{\rm h}21^{\rm m}31.91^{\rm s}$,
$\delta=+62^\circ38'39.2^{\prime\prime}$ (J2000), with a 90\% uncertainty
radius of $2.7^{\prime\prime}$. It is offset from the optical nucleus by \(1.85^{\prime\prime}\), well within its 90\% uncertainty radius, and is therefore consistent with a nuclear origin, as also shown in Figure~\ref{fig:fxt_loc}.

\subsection{\textit{Swift} UV/Optical Observations}

Ultraviolet data of EP~J062132.0+623837 from the Ultraviolet/Optical Telescope (UVOT; \citealt{Roming2005}) onboard the \textit{Swift} were obtained and processed using the standard \texttt{HEASoft} pipeline. Individual exposures were co-added with the \texttt{uvotimsum} task, and photometry was performed using \texttt{uvotsource}, adopting a circular source region with a radius of $5^{\prime\prime}$ and a concentric annular background region with inner and outer radii of $20^{\prime\prime}$ and $40^{\prime\prime}$, respectively.

\begin{figure*}
\centering
\includegraphics[width=\textwidth]{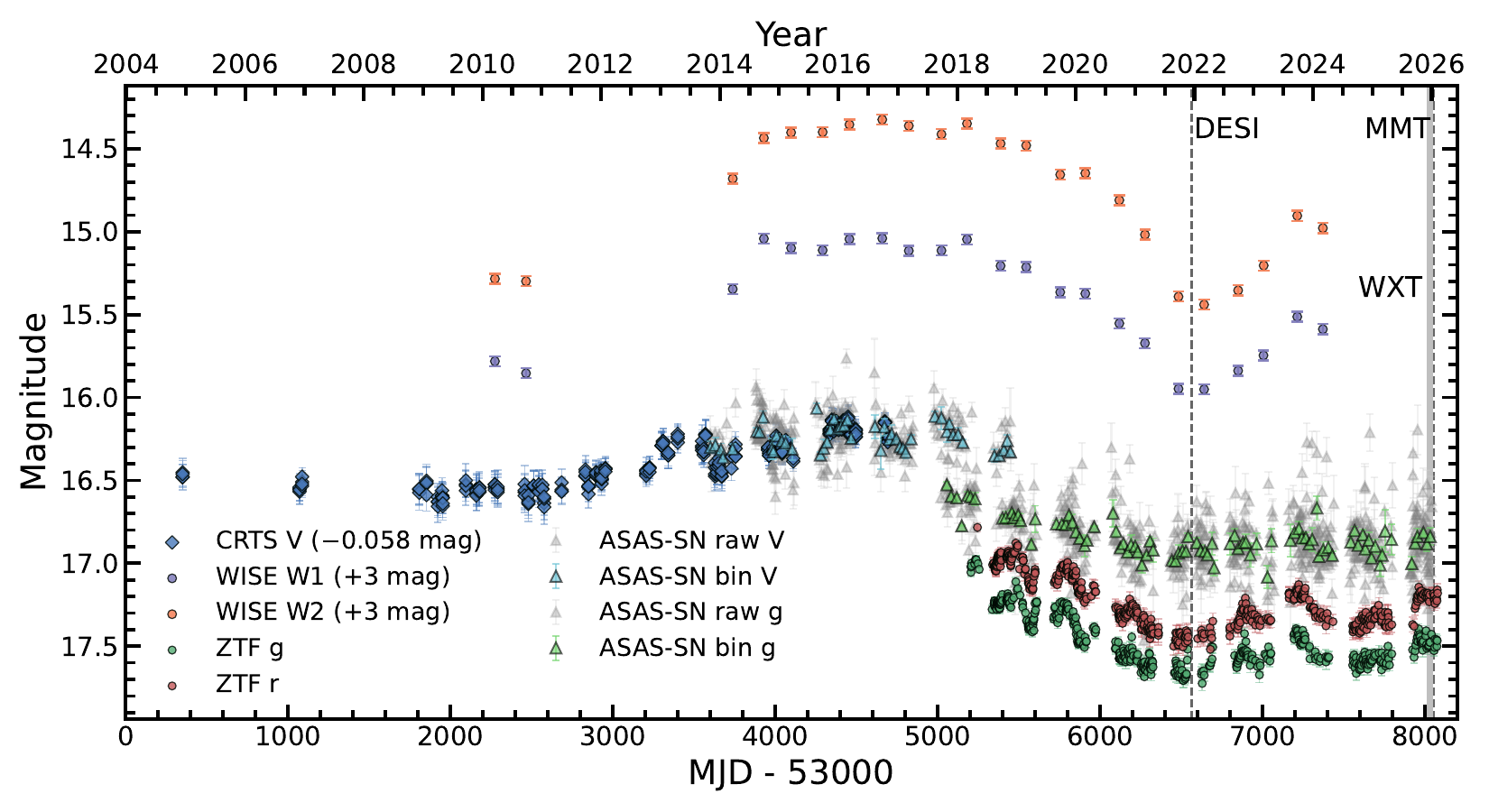}
\caption{
Long-term optical and MIR light curves of EP~J062132.0+623837. The optical data are compiled from CRTS, ASAS-SN, and ZTF, while the MIR data are from \textit{WISE}/\textit{NEOWISE}. The ASAS-SN measurements are shown both as individual raw points and as 20-day binned points. The CRTS light curve is shifted by $-0.058$ mag to align its zeropoint approximately with the ASAS-SN $V$-band light curve in the overlapping time range. For display purposes, the W1 and W2 light curves are shifted by $+3$ mag. The vertical dashed lines mark the epochs of the DESI and MMT spectroscopic observations, while the shaded region marks the approximate interval of significant EP/WXT detections. The optical measurements are otherwise shown in their native survey photometric systems, so small zeropoint differences between nominally similar bands across different surveys may remain. Because these surveys differ in aperture, calibration, and host-galaxy contribution, the optical light curves are intended mainly to illustrate the long-term variability pattern rather than to support precision cross-survey color comparison.
}
\label{fig:multiband_lc}
\end{figure*}

\subsection{Ground-based Optical Photometry}

To place the EP event in a broader temporal context, we compiled long-term optical photometry from the Catalina Real-Time Transient Survey (CRTS; \citealt{Drake2009}), the All-Sky Automated Survey for Supernovae (ASAS-SN; \citealt{Kochanek2017}), and the Zwicky Transient Facility \citep[ZTF; e.g.,][]{Bellm2019,Masci2019}. Together, these surveys provide an optical baseline of nearly two decades.

\subsubsection{ZTF Forced Photometry}

For ZTF, we used forced-photometry light curves from the ZTF Forced Photometry Service (ZFPS; \citealt{Masci2023ZFPS}), which provides difference-image forced photometry for ZTF survey data. Because ZFPS provides differential rather than directly plottable total-flux measurements, we reconstructed total-flux light curves by combining the differential fluxes with the reference-image photometry following the standard ZFPS framework. We applied the standard ZFPS quality cuts used in our reduction pipeline, performed robust clipping on the photometric zeropoints within each \((\mathrm{field}, \mathrm{ccdid}, \mathrm{qid}, \mathrm{filter})\) group, and retained only the dominant CCDID stream for the source to minimize offsets associated with different reference images. The resulting total fluxes were converted to AB magnitudes. We further corrected significant residual baseline offsets in the differential fluxes and renormalized the formal uncertainties when the light-curve scatter indicated that the pipeline errors were underestimated. Only measurements with positive total flux and total-flux signal-to-noise ratio \(>3\) were retained in the final \(g\)- and \(r\)-band light curves.

\subsubsection{ASAS-SN Photometry}

For ASAS-SN, we used the public \(V\)- and \(g\)-band light curves from the Sky Patrol database \citep{Kochanek2017}. The catalog times were converted from HJD to MJD. Measurements classified as non-detections were treated as upper limits, adopting the catalog limiting magnitude when available. We retained the \(V\)- and \(g\)-band data and removed obvious outliers.

\subsubsection{CRTS Photometry}

CRTS provides the earliest long-baseline optical coverage of the source \citep{Drake2009}. We used target-level CRTS photometric measurements, which primarily trace the variable point-source component rather than the total host-galaxy light. Obvious outliers were removed, and only measurements with finite photometric uncertainties were retained.

To improve the visual continuity of the long-term optical light curve, we applied a small empirical zeropoint offset of \(-0.058\) mag to the CRTS data relative to the ASAS-SN \(V\)-band measurements in their overlapping time range. This shift was used only for display consistency and is not intended as a strict cross-calibration between the two surveys.

\subsection{MIR Photometry}

We collected MIR photometry in the W1 and W2 bands from the NASA/IPAC IR Science Archive (IRSA), using single-exposure profile-fit measurements from the \textit{WISE} All-Sky and Post-Cryo releases and from the \textit{NEOWISE}-R release \citep[e.g.,][]{Wright2010,Mainzer2014}. These measurements were merged to construct unified W1 and W2 time series for the source.

Before binning, we removed exact duplicate measurements across the input files. We then grouped the single-exposure measurements into observing epochs, adopting a 3 day gap to separate adjacent visits. Within each epoch, we applied median absolute deviation (MAD)-based sigma clipping to reject obvious outliers and retained only epochs with at least five measurements. For each accepted epoch, we adopted the median observation time and median magnitude as the representative photometric point. The uncertainty of each binned point was estimated from the MAD-based scatter of the retained measurements, scaled to an equivalent Gaussian dispersion and divided by \(\sqrt{N}\).

Figure~\ref{fig:multiband_lc} presents the long-term optical and MIR light curves used in this work. The optical light curves are compiled from CRTS, ASAS-SN, and ZTF, while the MIR light curves are from \textit{WISE}/\textit{NEOWISE}. Except for the small empirical offset applied to the CRTS \(V\)-band data for display consistency, the optical measurements are shown in their native survey photometric systems. The vertical dashed lines mark the epochs of the DESI and MMT spectroscopy and provide a reference for the placement of the spectroscopic observations within the full long-term variability baseline.

\subsection{Radio}

We examined archival VLASS 3 GHz quicklook images at the position of EP~J062132.0+623837. Four epochs were inspected, and for each epoch we measured the local peak brightness, rms noise, and positional offset relative to the optical nucleus in small cutouts centered on the target position. In the 2017, 2020, and 2023 images, weak peaks are detected near the target position, with peak brightnesses of approximately 0.60, 1.00, and 0.74 mJy beam\(^{-1}\), respectively. A weaker and more marginal signal, with a peak brightness of about 0.54 mJy beam\(^{-1}\), is present in the 2025 image. In this work, we use the VLASS data as radio constraints on the nuclear counterpart.

\begin{figure*}
\centering
\includegraphics[width=\textwidth]{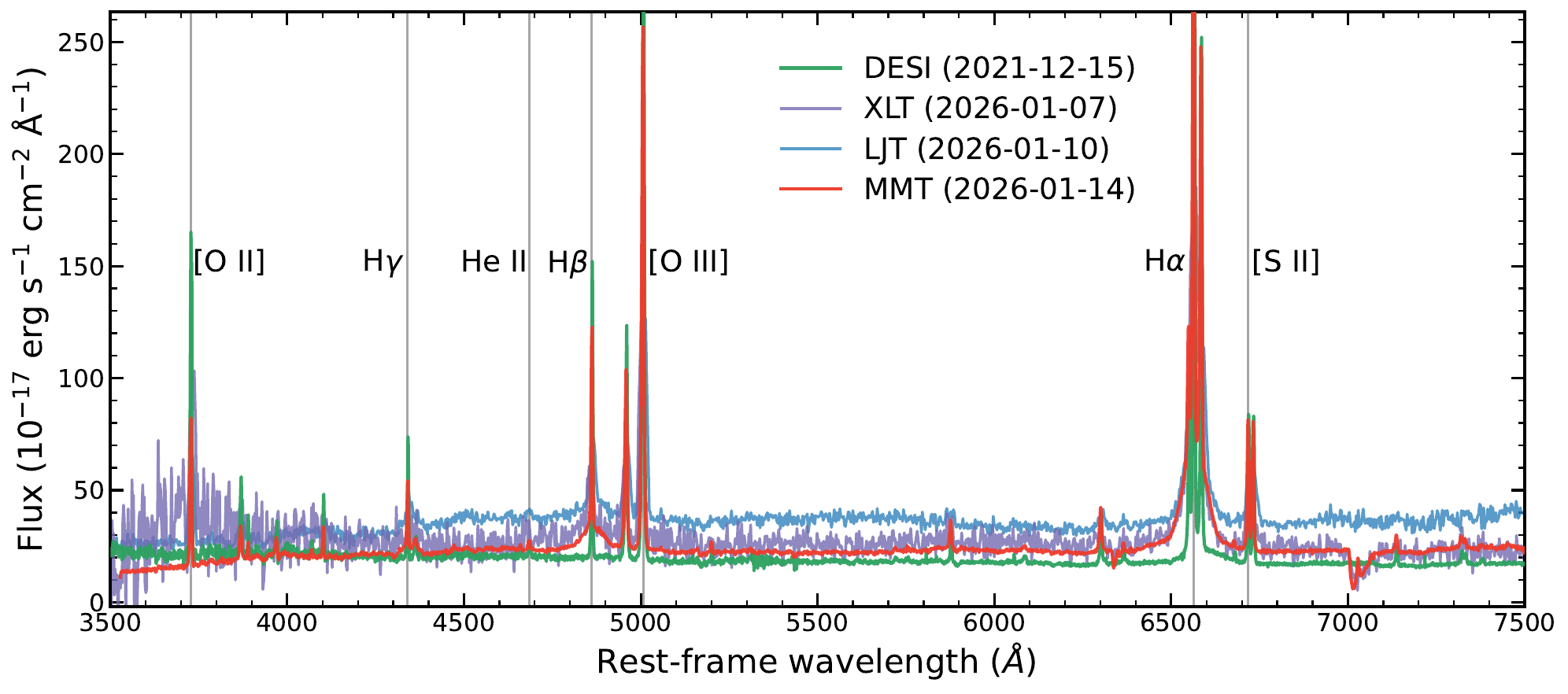}
\caption{
Rest-frame optical spectra of EP~J062132.0+623837. The green, purple, blue, and red curves show the DESI, XLT, LJT, and MMT spectra, respectively. The gray vertical lines mark the positions of the principal diagnostic emission lines. The corrected MMT spectrum is placed on the same relative flux scale as the DESI spectrum by matching the integrated flux of the [O\,\textsc{iii}]~\(\lambda5007\) emission line. The DESI and MMT spectra provide the main comparison discussed in the text, while the XLT and LJT spectra are shown for reference.
}
\label{fig:spec_compare}
\end{figure*}

\subsection{Optical Spectroscopy and Data Reduction}
\label{spec_obs}

Four optical spectra are available for EP~J062132.0+623837. The earliest was obtained by the Dark Energy Spectroscopic Instrument (DESI) on 2021 December 15, well before the EP discovery, and retrieved from DESI Data Release 1 \citep{DESIDR1}. The spectrum covers 3600.0--9755.2~\AA\ in the observed frame (3321.0--8999.3~\AA\ in the rest frame at z=0.084). Three additional spectra were acquired during the follow-up campaign: with the 2.16 m Xinglong Telescope (XLT) using the Beijing Faint Object Spectrograph and Camera (BFOSC; \citealt{Fan2016}) on 2026 January 7, the 2.4 m Lijiang Telescope (LJT) using the Yunnan Faint Object Spectrograph and Camera \citep[YFOSC; e.g.,][]{Fan2015YFOSC,Wang2019LJT} on 2026 January 10, and the 6.5 m Multiple Mirror Telescope (MMT) using Binospec \citep{Fabricant2019} on 2026 January 14. The XLT/BFOSC spectrum, obtained with the G4 grism, was reduced using an ASPIRED-based pipeline, including bias subtraction, flat-field correction, cosmic-ray removal, one-dimensional spectral extraction, FeAr-based wavelength calibration, flux calibration with a spectrophotometric standard star, and atmospheric-extinction correction. The final flux-calibrated spectrum covers 3850--9000~\AA\ in the observed frame. The LJT/YFOSC spectrum was reduced from the two-dimensional data using a locally adapted version of \texttt{PypeIt} \citep{Prochaska2020},  yielding a flux-calibrated spectrum spanning 3207.2--9329.4~\AA\ (2958.7--8606.5~\AA\ in the rest frame). The MMT/Binospec spectrum was provided by the observatory pipeline as a reduced, flux-calibrated product and covers 3824.3--9197.1~\AA\ (3527.9--8484.4~\AA\ in the rest frame). All four spectra include the key diagnostic regions around H$\beta$, [O~{\sc iii}], H$\alpha$, [N~{\sc ii}], and [S~{\sc ii}].


Since the three post-discovery spectra of EP~J062132.0+623837 were obtained within one week of each other, we adopt the MMT spectrum, which has the highest spectral resolution and S/N, for comparison with the archival DESI spectrum. However, the blue end of the MMT spectrum appears to be affected by uncertainties in the relative flux calibration, as evidenced by its discrepancy with the nearly contemporaneous XLT and LJT spectra. The standard Binospec pipeline provides relatively flux-calibrated spectra, for which slit losses are not accounted for \citep{Kansky2019}. We therefore applied an empirical large-scale correction to the blue-side continuum shape of the MMT spectrum, using the LJT spectrum as a reference. Specifically, after masking the major emission-line regions, we fitted the large-scale continuum flux ratio between the LJT and MMT spectra with a low-order polynomial and applied the resulting wavelength-dependent correction to the MMT spectrum. The corrected MMT spectrum was then scaled to the DESI spectrum by matching the integrated flux of the [O\,\textsc{iii}]~\(\lambda5007\) emission line under the assumption that the [O\,\textsc{iii}] emission did not vary significantly between the two epochs. All subsequent spectral comparisons are based on the DESI spectrum and the processed MMT spectrum placed on a common relative flux scale in this way. Figure~\ref{fig:spec_compare} presents the DESI, XLT, LJT, and the corrected MMT spectra.

\section{Results}\label{sec:3}
\subsection{Multiwavelength Light Curves}

\begin{figure}
\centering
\includegraphics[width=0.45\textwidth]{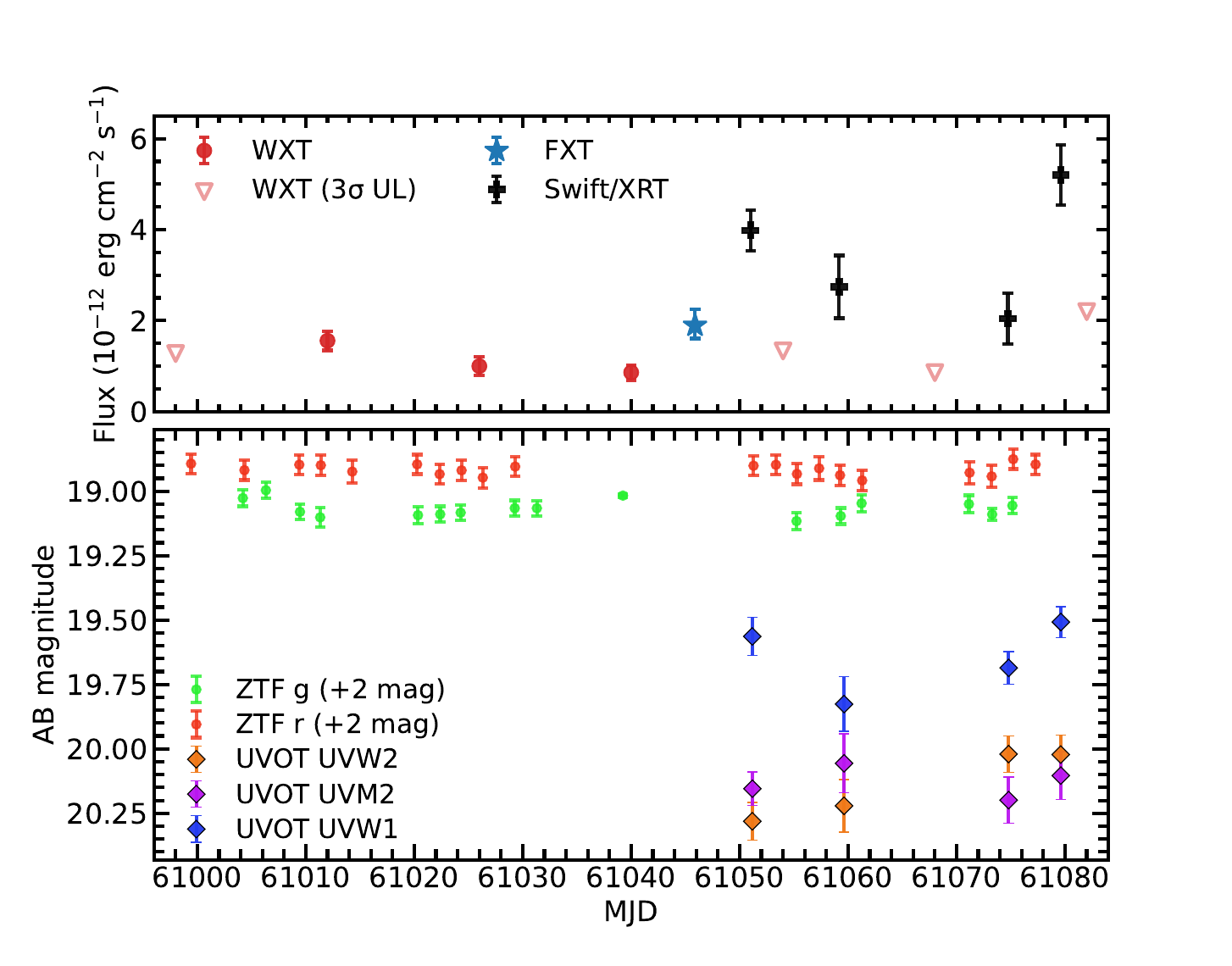}
\caption{
Recent optical, UV, and X-ray light curves of EP~J062132.0+623837 around the EP discovery. The \emph{bottom} panel shows ZTF \(g\)- and \(r\)-band photometry together with \textit{Swift}/UVOT measurements in the UVW2, UVM2, and UVW1 bands. All optical and UV magnitudes in the bottom panel have been corrected for Galactic extinction and are shown in the AB system. For display clarity, the ZTF \(g\)- and \(r\)-band magnitudes are shifted by \(+2\) mag. No host-galaxy subtraction has been applied. The \emph{top} panel shows X-ray fluxes from WXT (0.5--4.0 keV), FXT (0.5--10 keV), and \textit{Swift}/XRT (0.3--10 keV). WXT measurements with detection significance \(\ge 5\) are shown as detections, while bins with significance \(<5\) are shown as \(3\sigma\) upper limits. Because the X-ray measurements are derived in different instrumental energy bands, the X-ray light curve is intended mainly to illustrate the temporal sequence of the post-discovery activity. Error bars denote \(1\sigma\) uncertainties.
}
\label{fig:recent_lc}
\end{figure}

The photometry compiled from CRTS, ASAS-SN, and ZTF provides a long-term optical light curve spanning approximately two decades (Figure~\ref{fig:multiband_lc}). To place the heterogeneous optical data sets on a common variability scale, we intercalibrated the CRTS and ASAS-SN \(V\)-band and ZTF \(g\)-band light curves using PyCALI \citep{Li2014}, allowing for relative multiplicative and additive offsets between data sets. The calibrated measurements were then combined into a single long-term optical continuum light curve and placed on a common flux-density scale anchored to the ZTF \(g\)-band photometry. It reveals a pronounced brightening beginning around 2010, followed by a gradual decline starting around 2018 and substantial variability thereafter. The ZTF data offer the clearest view of the recent variability, exhibiting peak-to-trough amplitudes of approximately ($\Delta g \approx 0.75$) mag and ($\Delta r \approx 0.7$) mag. The MIR light curves from \textit{WISE}/\textit{NEOWISE} broadly follow the overall optical variability trend, albeit with a much lower cadence. Over the available WISE/NEOWISE baseline, the source shows large-amplitude MIR variability, with peak-to-trough changes of \(\Delta W1 \approx 0.91\) mag and \(\Delta W2 \approx 1.12\) mag after applying the same epoch-binning procedure as used in Figure~\ref{fig:multiband_lc}. This behavior is consistent with a delayed and smoothed response of dust-reprocessed emission to variations in the accretion-powered optical continuum \citep[e.g.,][]{Netzer2015,Sheng2017}. To more clearly illustrate the short-term variability before and after the EP discovery, we also present the recent optical, UV, and X-ray photometric data in Figure~\ref{fig:recent_lc}, which shows much smaller variability amplitudes in all bands. Taken together, the historical and recent light curves suggest that EP~J062132.0+623837 had been intensely variable for many years prior to its X-ray discovery by EP, indicating that the EP detection is part of a long-lived activity rather than an isolated nuclear flare.



\begin{figure}
\centering
\includegraphics[width=0.45\textwidth]{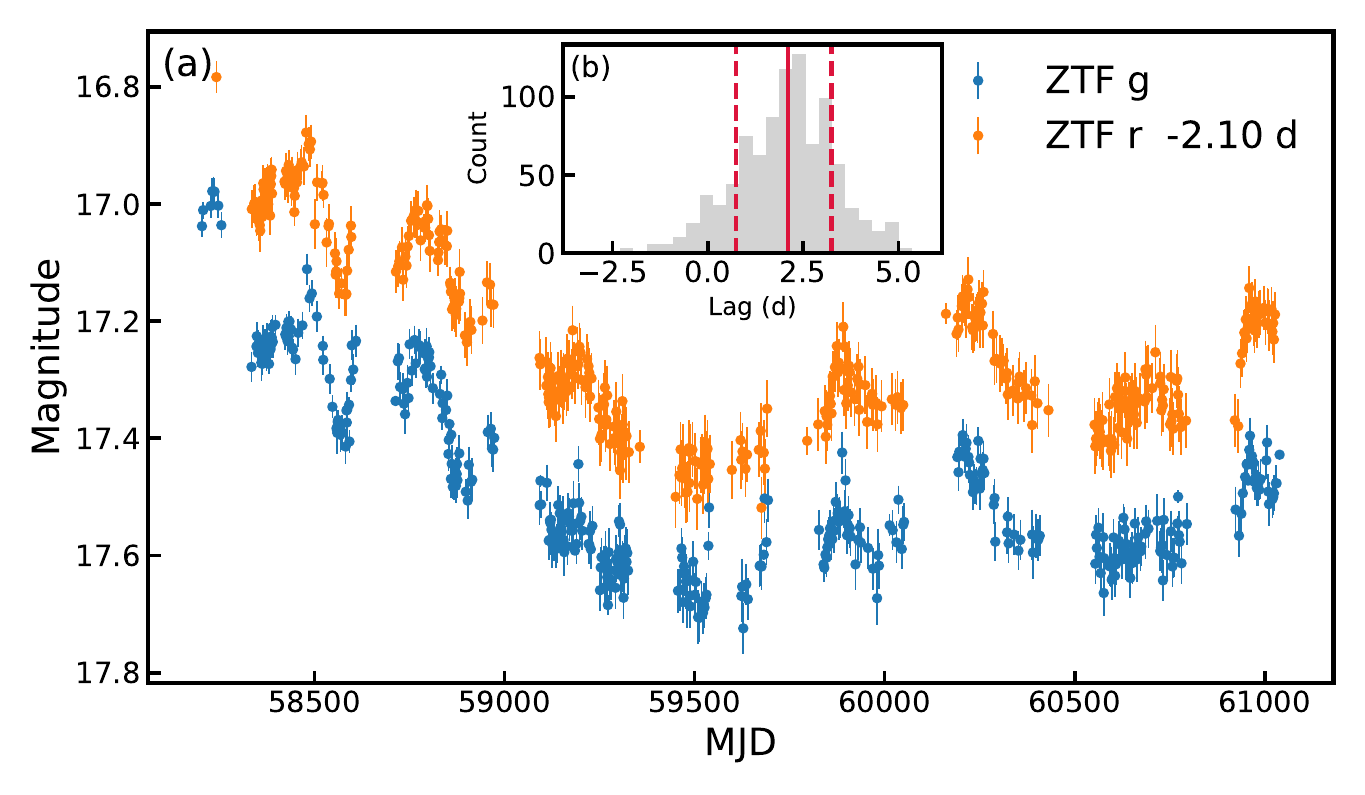}
\caption{
Lag analysis of the recent ZTF optical variability of EP~J062132.0+623837. 
Panel (a) compares the ZTF \(g\)-band light curve with the ZTF \(r\)-band light curve shifted by \(-2.10\) d, illustrating the approximate alignment obtained for the preferred lag. 
Panel (b) shows the FR/RSS centroid distribution derived from the ICCF analysis. 
The red solid line marks the median lag, and the black dashed lines mark the 16th and 84th percentiles of the centroid distribution. 
The resulting lag is \(\tau_{r-g} = 2.10\) d, with a 16th--84th percentile range of 0.74--3.25 d, in the sense that the \(r\)-band variations lag behind the \(g\)-band variations.
}
\label{fig:grlag}
\end{figure}

The long-term highly variable multiband light curves further enable an investigation of the inter-band timing behavior. We analyzed the recent ZTF \(g\)- and \(r\)-band light curves using the interpolated cross-correlation function (ICCF) implemented in PyCCF \citep{Sun2018PyCCF}, and estimated the lag uncertainties with the standard flux-randomization/random-subset-selection (FR/RSS) Monte Carlo procedure \citep{Peterson1998}. We adopt the convention that \(\tau>0\) indicates that the \(r\)-band variations lag those in the \(g\) band. The ICCF analysis yields a lag of \(\tau_{r-g}=2.10\) d, with a 16th--84th percentile range of 0.74--3.25 d, in the sense that the \(r\)-band variations lag those in the \(g\) band (Figure~\ref{fig:grlag}). The FR/RSS centroid distribution also favors a positive lag of a few days. This lag is broadly consistent with wavelength-dependent optical continuum variability, whereby the bluer band responds slightly earlier than the redder band. However, given the observational cadence and the relatively small amplitude of the short-timescale variability, we consider this result to be indicative rather than definitive and do not interpret it as a precise reverberation measurement.

For the optical--MIR timing analysis, we used the full PyCALI-calibrated optical continuum together with the epoch-binned W1 and W2 light curves. Following \citet{Lyu2019}, we modeled the optical variability as a damped random walk (DRW) and used the resulting continuous reconstruction as the driving light curve. For each trial delay \(\Delta t\), the optical model was smoothed with a top-hat window of width \(\tau_{\rm W}=\Delta t/2\), capped at 200~d, and shifted, scaled, and offset to fit W1 and W2 independently:
\begin{equation}
F_{\rm MIR}(t)
=
{\rm AMP}\,
\left\langle F_{\rm opt}(t-\Delta t)\right\rangle_{\tau_{\rm W}}
+
F_{\rm const}.
\end{equation}
The preferred delay was determined from the minimum-\(\chi^{2}\) solution.

\begin{figure}
\centering
\includegraphics[width=0.45\textwidth]{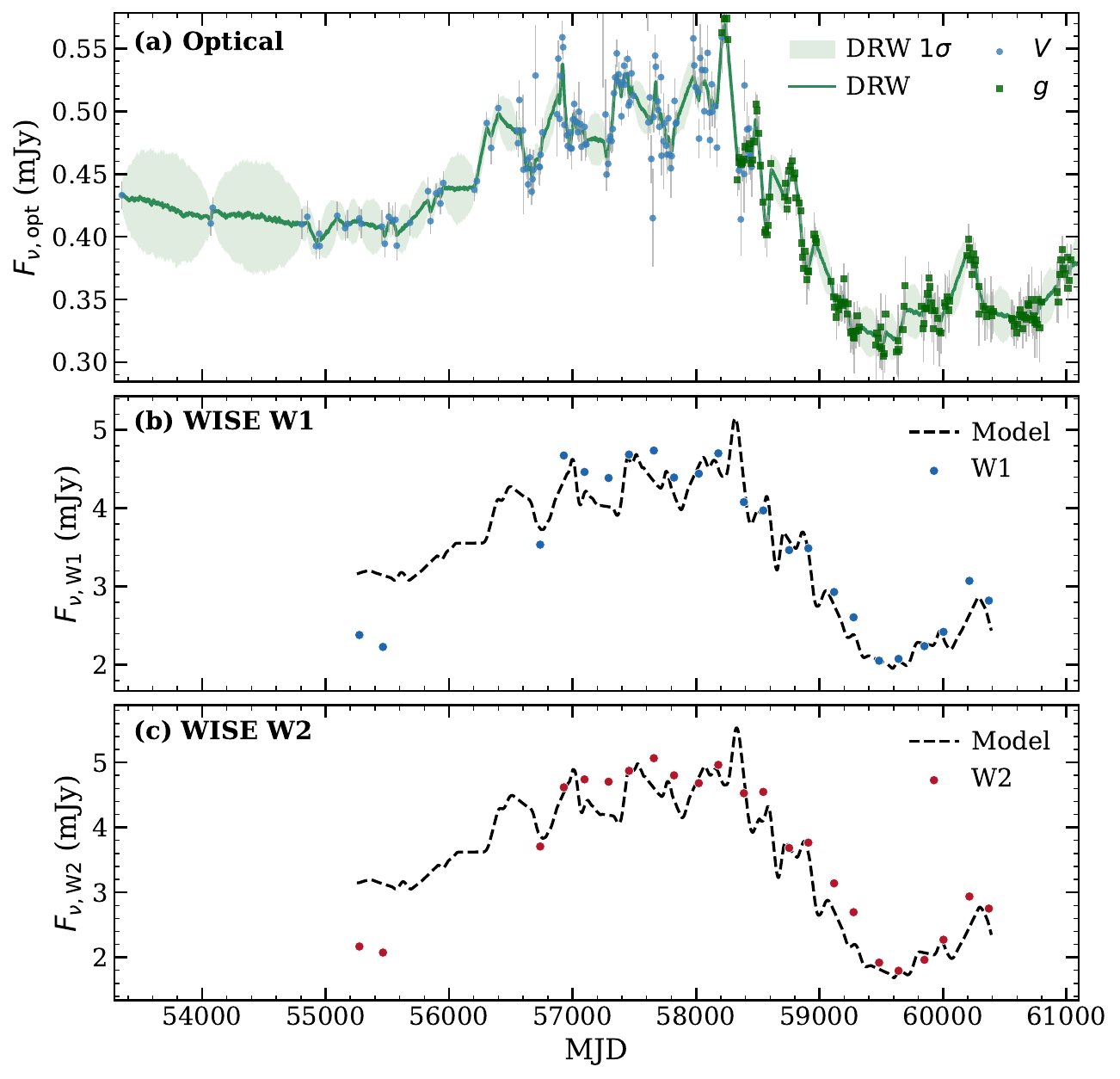}
\caption{Optical--MIR reverberation analysis of EP~J062132.0+623837 following the approach of \citet{Lyu2019}. Panel (a) shows the PyCALI-intercalibrated optical continuum, with the \(V\)-band and ZTF \(g\)-band measurements shown separately. For clarity, the optical measurements are displayed in 10-day bins, whereas the DRW reconstruction was derived from the full calibrated optical light curve. The solid green curve shows the median DRW reconstruction and the shaded region denotes its \(1\sigma\) interval. Panels (b) and (c) show the epoch-binned \textit{WISE}/\textit{NEOWISE} W1 and W2 light curves, respectively, together with the best-fitting delayed and smoothed optical models. The corresponding observed-frame delays are
\(\Delta t_{\rm W1}=85.5\pm31.4\)~d and
\(\Delta t_{\rm W2}=97.5\pm63.5\)~d.
}
\label{fig:optmir}
\end{figure}

Uncertainties associated with the optical interpolation were propagated using 1000 conditional DRW realizations, with the MIR fit repeated for each realization. Following \(3\sigma\) clipping, the dispersion of the recovered delays was combined in quadrature with the formal MIR fitting uncertainty. We obtain observed-frame delays of \(\Delta t_{\rm W1}=85.5\pm31.4\)~d and \(\Delta t_{\rm W2}=97.5\pm63.5\)~d, corresponding to rest-frame delays of \(78.9\pm29.0\)~d and \(89.9\pm58.6\)~d, respectively (Figure~\ref{fig:optmir}).

As an independent check, we computed ICCFs between the 10-day-binned optical continuum and the W1 and W2 light curves, with lag uncertainties estimated from 5000 FR/RSS realizations. Positive lag denotes MIR variations lagging the optical continuum. The dominant ICCF peaks occur at \(\sim106\)~d in W1 and \(\sim126\)~d in W2. The corresponding centroid distributions give
\(\tau_{\rm cent}=91.7^{+27.0}_{-59.0}\)~d for W1 and \(120.2^{+75.0}_{-24.3}\)~d for W2. Although the FR/RSS distributions are non-Gaussian and show secondary structure, the two complementary methods both favor a positive optical--MIR delay on a characteristic timescale of \(\sim10^{2}\)~d.



\subsection{X-Ray Spectral Properties}

The post-discovery X-ray spectral properties of EP~J062132.0+623837 are derived from the FXT and \textit{Swift}/XRT follow-up data and are summarized in Table~\ref{tab:xray}. 

\begin{table*}[t]
\centering
\caption{Summary of the X-ray observations and spectral properties of EP~J062132.0+623837}
\label{tab:xray}
\small
\setlength{\tabcolsep}{5pt}
\begin{tabular}{lccccccc}
\hline
Instrument & MJD & Exposure & Band & Observed flux & Observed luminosity & $\Gamma$ & Fit statistic \\
 &  & (ks) & (keV) & ($10^{-12}\ \mathrm{erg\ cm^{-2}\ s^{-1}}$) & ($10^{43}\ \mathrm{erg\ s^{-1}}$) &  &  \\
\hline
EP/FXT & 61045.886 & 8.54 & 0.5--10 & $1.79^{+0.33}_{-0.30}$ & $3.14^{+0.58}_{-0.53}$ & $1.65 \pm 0.22$ & $21.09/24$ \\

Swift/XRT & 61051.003 & 3.12 & 0.3--10 & $3.98 \pm 0.44$ & $6.99 \pm 0.77$ & \multirow{4}{*}{$1.50 \pm 0.14$} & \multirow{4}{*}{$20.79/18$} \\
Swift/XRT & 61059.174 & 2.40 & 0.3--10 & $2.74 \pm 0.69$ & $4.81 \pm 1.21$ &  &  \\
Swift/XRT & 61074.731 & 2.04 & 0.3--10 & $2.05 \pm 0.56$ & $3.60 \pm 0.98$ &  &  \\
Swift/XRT & 61079.610 & 1.80 & 0.3--10 & $5.20 \pm 0.67$ & $9.13 \pm 1.18$ &  &  \\
\hline
\end{tabular}

\vspace{1mm}
\begin{minipage}{0.97\linewidth}
\footnotesize
Notes. Spectral parameters are derived using the XSPEC model \texttt{tbabs*zashift*powerlaw}, with the Galactic absorption column density and source redshift fixed as described in the text. Fluxes and luminosities are observed values in the fitted band and are not corrected for absorption. For \textit{Swift}/XRT, the photon index and fit statistic are derived from the stacked post-discovery spectrum, while the four individual epochs are listed to show the single-epoch flux evolution. The individual-epoch fluxes were calculated from the background-subtracted count rates using the 0.3--10 keV count-rate-to-flux conversion factor derived from the best-fitting stacked spectrum. The fit statistic is reported as statistic/dof, using C-statistic for the lower-count FXT spectrum and \(\chi^2\) for the grouped stacked \textit{Swift}/XRT spectrum.
\end{minipage}
\end{table*}

The FXT follow-up observation localized the X-ray source to the nuclear region of 2MASX J06213207+6238390 (Figure~\ref{fig:fxt_loc}). We fitted the FXT spectrum in \texttt{xspec v12.15.1} \citep{Arnaud1996} with an absorbed redshifted power-law model, fixing the Galactic absorption column density at \(N_{\rm H} = 8.66\times10^{20}\ \mathrm{cm^{-2}}\) based on the HI4PI map \citep{HI4PI2016} and the source redshift at \(z = 0.084\). The FXT spectrum is adequately described by this model, with a photon index of \(\Gamma = 1.65 \pm 0.22\), C-stat/dof \(=21.09/24\), and an observed 0.5--10 keV flux of \((1.79^{+0.33}_{-0.30})\times10^{-12}\ \mathrm{erg\ cm^{-2}\ s^{-1}}\).  No additional soft component is required by the fit.

We obtained four \textit{Swift}/XRT follow-up observations between 2026 January 11 and 2026 February 8. The individual XRT fluxes remain in the range \(2.05\text{--}5.20\times10^{-12}\ \mathrm{erg\ cm^{-2}\ s^{-1}}\) in the 0.3--10 keV band, indicating that the source stayed X-ray active throughout the initial follow-up interval (Figure~\ref{fig:recent_lc}). We stacked all four observations and fitted the XRT spectrum in the same manner as for FXT, adopting the same absorbed redshifted power-law model and the same fixed Galactic absorption column. The resulting XRT spectrum yields a photon index of \(\Gamma = 1.50 \pm 0.14\), with \(\chi^2\)/dof \(=20.79/18\), consistent with the FXT result. The best-fitting stacked spectrum was used to derive a 0.3--10 keV count-rate-to-flux conversion factor of \(1.004\times10^{-10}\ \mathrm{erg\ cm^{-2}\ count^{-1}}\), which was then applied to the background-subtracted count rate of each individual observation to obtain the single-epoch fluxes listed in Table~\ref{tab:xray}. Although small residuals remain, the stacked \textit{Swift}/XRT spectrum does not show statistically compelling large-scale structure, and we therefore do not introduce a more complex spectral model.

\begin{figure}
\centering
\includegraphics[width=0.45\textwidth]{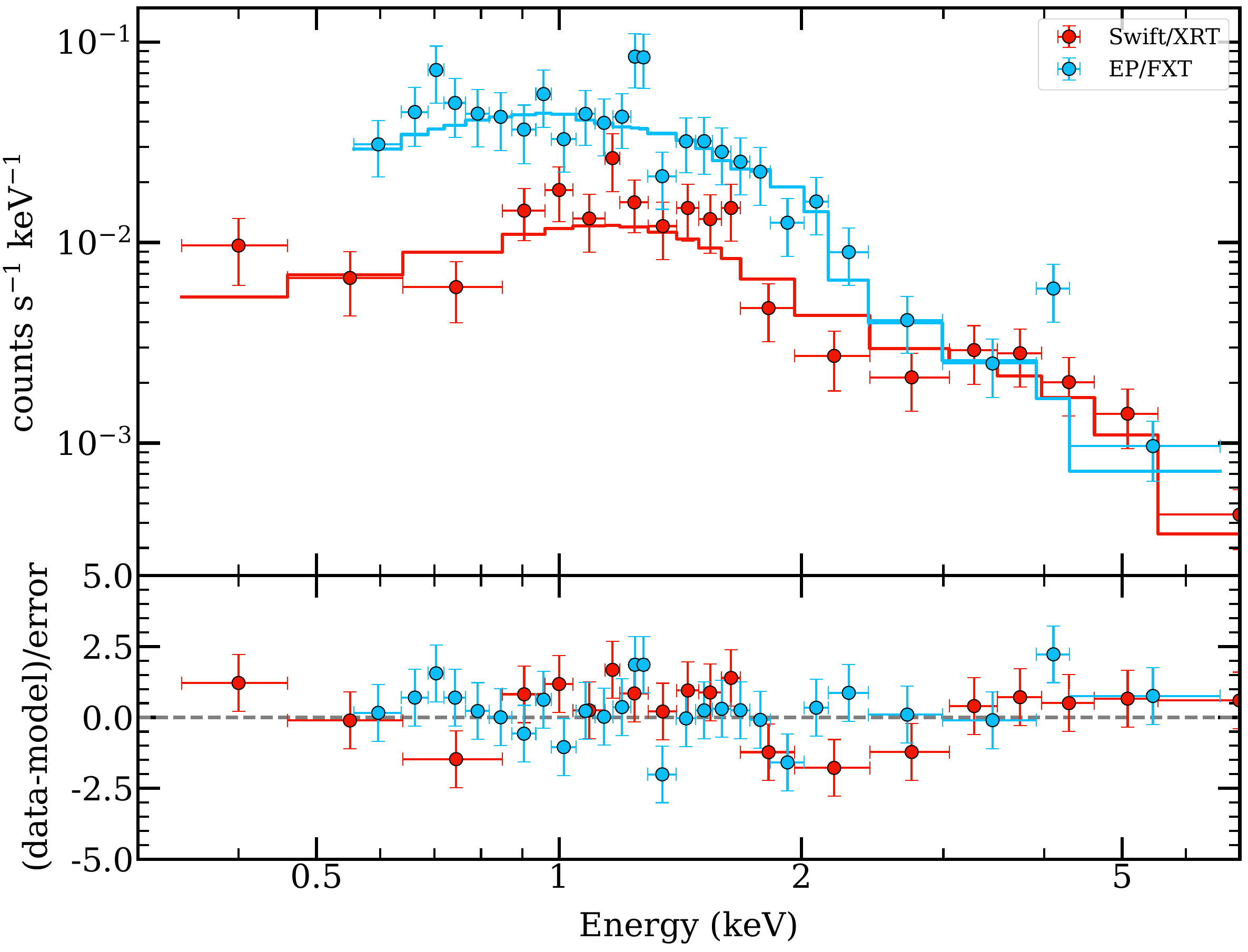}
\caption{
X-ray spectra of EP~J062132.0+623837. The FXT spectrum (blue) and the combined \textit{Swift}/XRT spectrum (red) are shown together in the upper panel, along with their best-fit XSPEC models (\texttt{tbabs*zashift*powerlaw}). The FXT spectrum is fitted over 0.5--10 keV, while the combined \textit{Swift}/XRT spectrum is fitted over 0.3--10 keV. The lower panel shows the residuals relative to the best-fit models, expressed as \((\mathrm{data}-\mathrm{model})/\mathrm{error}\). 
}
\label{fig:xray_spec}
\end{figure}

Figure~\ref{fig:xray_spec} shows the FXT and combined \textit{Swift}/XRT spectra together with their best-fit absorbed redshifted power-law models. The FXT and \textit{Swift}/XRT spectra are mutually consistent and show that the source remained X-ray active during the initial post-discovery follow-up interval.

\subsection{Optical Spectral Properties and Black Hole Mass Estimate}
\label{spec_analysis}

\begin{figure*}
\centering
\includegraphics[width=0.8\textwidth]{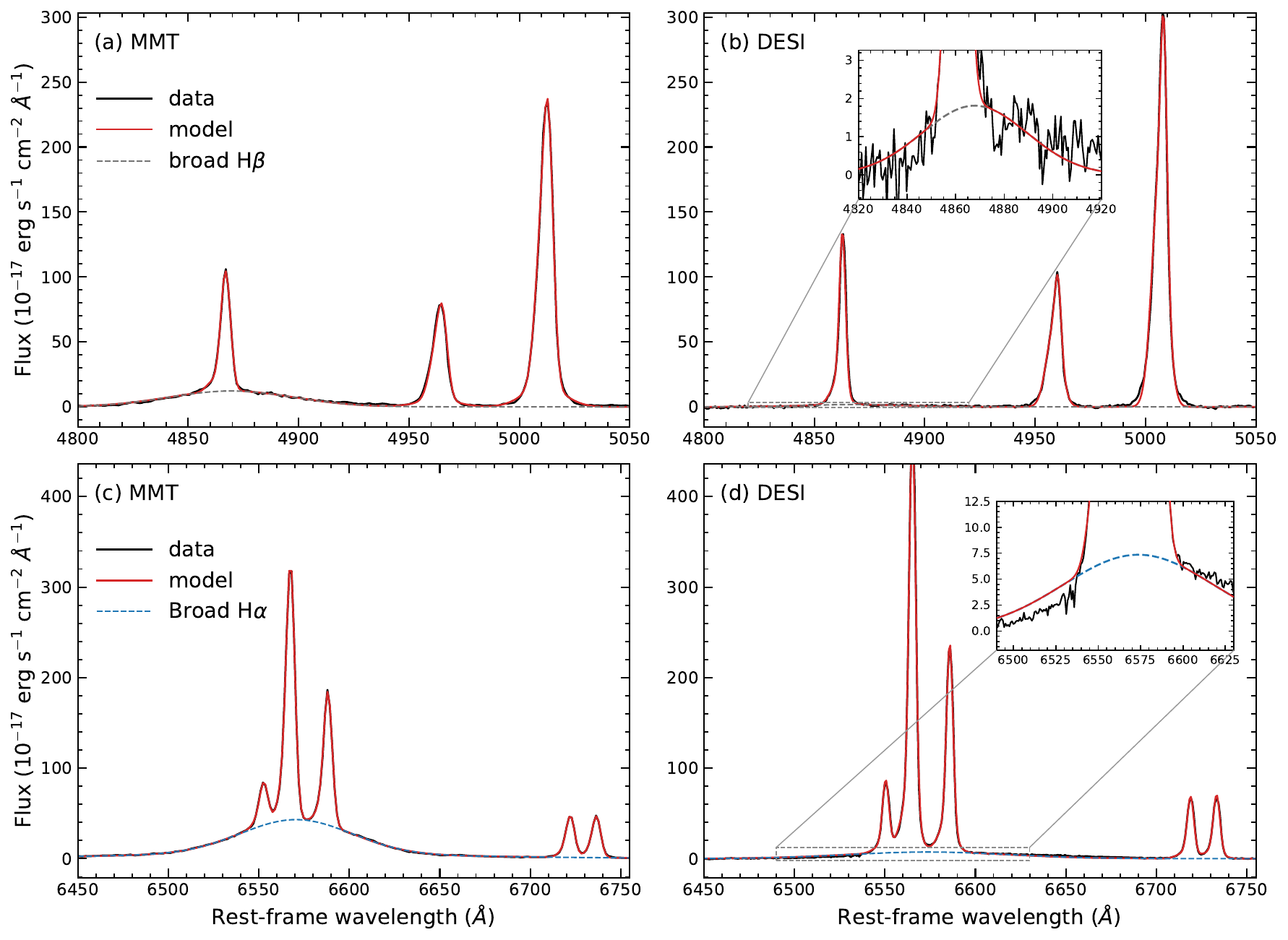}
\caption{
Comparison of the Balmer-line regions in the pre-discovery DESI spectrum and the post-discovery MMT spectrum of EP~J062132.0+623837. Panels (a) and (b) show the host-subtracted H$\beta$+[O\,\textsc{iii}] region for the MMT and DESI spectra, respectively, while panels (c) and (d) show the corresponding H$\alpha$ regions. In all panels, the black curve shows the host-subtracted spectrum and the red curve shows the total model. In panels (a) and (b), the gray dashed curve marks the broad H$\beta$ component. In panels (c) and (d), the blue dashed curve marks the summed broad H$\alpha$ component. The insets in panels (b) and (d) show enlarged views of the DESI H$\beta$ and H$\alpha$ regions, respectively, to highlight the weak broad-line components. The DESI spectrum shows broad H$\alpha$ and a weak broad H$\beta$ component, while the MMT spectrum shows a substantially stronger broad H$\beta$ component together with a stronger and more structured broad H$\alpha$ profile. The adopted reduced $\chi^2$ values are 6.69 and 11.67 for the DESI and MMT fits in the H$\beta$+[O\,\textsc{iii}] region, respectively, and 7.71 and 2.34 for the DESI and MMT fits in the H$\alpha$ region, respectively.
}
\label{fig:balmer_compare}
\end{figure*}

We use the DESI and MMT spectra to study the spectral evolution before and after the EP discovery of EP~J062132.0+623837. The most notable change is the enhancement of broad H\(\alpha\) and H\(\beta\) emission, as shown in Figure~\ref{fig:balmer_compare}. To quantify these spectral variations, we modeled and subtracted the stellar continuum using \texttt{pPXF} \citep{Cappellari2017} with stellar templates from the E-MILES library \citep{Vazdekis2016}, and then fitted the emission-line complexes in the continuum-subtracted spectra. The narrow-line profiles were modeled with two Gaussian components per line to empirically capture the observed asymmetry and profile complexity. In practice, a single narrow Gaussian leaves significant systematic residuals in several strong narrow lines, particularly in the [O\,\textsc{iii}] and H\(\alpha\)+[N\,\textsc{ii}] regions, whereas the two-component model provides a more satisfactory empirical description. We emphasize that we do not assign separate physical meaning to the individual narrow components.

For the DESI H\(\beta\)+[O\,\textsc{iii}] region, we adopted a model consisting of two [O\,\textsc{iii}] components, two narrow H\(\beta\) components tied to the narrow-line kinematics, and one broad H\(\beta\) component. Fits without a broad H$\beta$ component leave noticeably larger residuals around the H$\beta$ profile. Because the velocity centroid of the broad H\(\beta\) component is not well constrained when left free, we tie its velocity offset to that of the better-constrained broad H\(\alpha\) component in the adopted decomposition. To assess whether the weak broad H$\beta$ component is statistically favored in the DESI spectrum, we compared otherwise identical models with and without this component, while keeping its velocity offset tied to that of the better-constrained broad H$\alpha$ line. The inclusion of broad H$\beta$ improves the reduced $\chi^2$ from 7.98 to 6.69, with $\Delta\chi^2=561.3$ for two additional free parameters. The corresponding F-test gives $F=41.9$ and $p=2.5\times10^{-17}$, while the model comparison also yields $\Delta{\rm AIC}=557.3$ and $\Delta{\rm BIC}=549.2$ in favor of the broad-component model. We therefore regard the broad H$\beta$ component as statistically favored, while treating the F-test only as a heuristic indicator rather than a standalone model-selection criterion \citep{Protassov2002}. For the MMT H\(\beta\)+[O\,\textsc{iii}] region, the decomposition is more challenging because the blue side has been corrected using a large-scale empirical shape correction (see Section~\ref{spec_obs}) and the [O\,\textsc{iii}] profile appears more complex. We therefore adopt a more flexible local model for the MMT H\(\beta\)+[O\,\textsc{iii}] region. The resulting fit yields a relatively large reduced \(\chi^2 = 11.67\), indicating that modest residual mismatches remain between the model and the data. We attribute these primarily to the combination of very high S/N, residual uncertainties in the empirical blue-side correction, and the intrinsically complex [O\,\textsc{iii}] profile. As in the DESI case, the broad H\(\beta\) velocity offset is tied to that of the better-constrained broad H\(\alpha\) component to stabilize the local decomposition.

For the H\(\alpha\) complex, after subtraction of the host-galaxy continuum, we also used two Gaussian components per narrow line to account for the narrow H\(\alpha\) emission. For the broad H\(\alpha\) emission, a single broad H\(\alpha\) component already provides an adequate description of the broad-line profile, whereas the MMT broad H\(\alpha\) profile is better represented by a two-component broad-line model.

Figure~\ref{fig:balmer_compare} compares the host-subtracted wavelength regions of the H\(\beta\)+[O\,\textsc{iii}] and H\(\alpha\) complexes. In the DESI spectrum, broad H\(\alpha\) is clearly present, while broad H\(\beta\) is weak but detectable. The enlarged views in panels (b) and (d) highlight these weak broad components in the DESI spectrum. In the MMT spectrum, both the broad H\(\alpha\) and  broad H\(\beta\) become much stronger. This qualitative change suggests a potential transition in AGN subtype. To make this subtype change more explicit, we adopt the classical intermediate-type Seyfert classification scheme of \citet{Osterbrock1977,Osterbrock1981} and use the total H\(\beta\)-to-[O\,\textsc{iii}] flux ratio as an empirical quantitative indicator, following \citet{Cohen1986,Winkler1992}. From the adopted local decompositions listed in Appendix~\ref{app:linemeas}, we obtain \(R_{\rm tot}=0.324^{+0.004}_{-0.004}\) for the DESI spectrum and \(R_{\rm tot}=0.700^{+0.005}_{-0.005}\) for the MMT spectrum, where \(R_{\rm tot}\equiv F({\rm H}\beta_{\rm tot})/F([\mathrm{O\,III}]\,\lambda5007)\). The fitted broad H\(\beta\) flux also increases substantially between the two epochs. In this framework, the DESI spectrum is classified as Type~1.8, while the MMT spectrum is classified as Type~1.5. The enhanced broad H\(\beta\) emission in the MMT spectrum is the primary driver of this subtype change. Detailed emission-line measurements are provided in Appendix~\ref{app:linemeas}.

From the measurements in Appendix~\ref{app:linemeas}, the broad H\(\alpha\) flux increases from \(837.24^{+9.80}_{-10.09}\times10^{-17}\ {\rm erg\ s^{-1}\ cm^{-2}}\) in the DESI spectrum to \(4090.46^{+12.67}_{-14.82}\times10^{-17}\ {\rm erg\ s^{-1}\ cm^{-2}}\) in the MMT spectrum, corresponding to luminosities of \(\log L_{\mathrm{H}\alpha,\mathrm{DESI}} = 41.166^{+0.005}_{-0.005}\) and \(\log L_{\mathrm{H}\alpha,\mathrm{MMT}} = 41.855^{+0.001}_{-0.002}\), respectively. The quoted luminosity uncertainties were propagated from the emission-line flux uncertainties and the DESI redshift uncertainty, with the latter contributing negligibly. Meanwhile, the FWHM of broad H\(\alpha\) decreases from \(4786.0^{+80.8}_{-82.2}\ \mathrm{km\ s^{-1}}\) to \(3524.0^{+14.8}_{-12.7}\ \mathrm{km\ s^{-1}}\), consistent with the AGN ``breathing'' effect \citep[e.g.,][]{Peterson2004,Bentz2009}.
We estimate the black hole mass using the H\(\alpha\)-based single-epoch virial relation of \citet{GreeneHo2005},
\begin{equation}
M_{\rm BH} = 2.0\times10^{6}
\left(\frac{L_{\mathrm{H}\alpha}}{10^{42}\ \mathrm{erg\ s^{-1}}}\right)^{0.55}
\left(\frac{\mathrm{FWHM}_{\mathrm{H}\alpha}}{10^{3}\ \mathrm{km\ s^{-1}}}\right)^{2.06}
M_\odot .
\end{equation}
This yields \(\log(M_{\rm BH}/M_\odot)=7.24\pm0.02\) for the DESI spectrum and \(7.35\pm0.01\) for the MMT spectrum, where the quoted uncertainties reflect the propagated statistical uncertainties in the broad H$\alpha$ luminosity and FWHM measurements. These formal uncertainties are much smaller than the intrinsic
scatter of the single-epoch virial mass estimator, which is
approximately \(0.5\) dex \citep{GreeneHo2005}. These estimates indicate a characteristic black hole mass of $\log(M_{\rm BH}/M_\odot)\approx7.3$ for EP~J062132.0+623837.


Taken together, these measurements show that the source remained in a broad-line state across both epochs, but that its Balmer-line properties changed substantially between the pre-discovery and post-discovery observations. The much stronger broad H$\beta$ component in the MMT spectrum, together with the contemporaneous increase in broad H$\alpha$ strength, is consistent with a changing-look transition from Type~1.8 to Type~1.5 within the broader class of intermediate-type transitions \citep[e.g.,][]{Denney2014,Li2022,RicciTrakhtenbrot2023}.

\section{Discussion}\label{sec:4}

\subsection{EP~J062132.0+623837 in the context of changing-look AGNs}

The optical spectral evolution of EP~J062132.0+623837, marked by the enhancement of the Balmer emission lines between the pre-discovery DESI spectrum and the post-discovery MMT spectrum, reveals a transition from a Type~1.8 to a Type~1.5 AGN (see Section~\ref{spec_analysis}). This intermediate-type transition is characteristic of the broader population of changing-look AGNs~\citep[e.g.,][]{Denney2014,Li2022,Guo2024DESI}. The primary driver of changing-look AGNs is believed to be a dramatic change in the accretion rate rather than variations in line-of-sight obscuration, as supported by observations of delayed infrared emission \citep[e.g.,][]{Sheng2017,Lyu2022,RicciTrakhtenbrot2023} and optical polarization measurements \citep{Hutsemekers2017}. This also appears to be the case for EP~J062132.0+623837, whose long-term optical and MIR light curves show broadly similar variability patterns, while the lag analyses favor a delayed MIR response on a timescale of order \(10^2\) d, as expected from dust reprocessing. However, its MIR color evolution is less extreme than in the MIR-selected turn-off CLAGNs discussed by \citet{Sheng2020}. In that work, typical turn-off CLAGNs show a transition from AGN-like MIR color (\(W1-W2>0.8\)) to galaxy-like color (\(W1-W2<0.5\)). By contrast, the \(W1-W2\) color of EP~J062132.0+623837 varies only within \(\sim 0.48\)--0.75. The source therefore shows substantial MIR variability, but not the full AGN-like to galaxy-like MIR color transition seen in more extreme MIR-selected cases.


Using the broad H$\alpha$ luminosities measured from the optical spectra and the empirical relation \(\displaystyle L_{\mathrm{H}\alpha} =(5.25\pm0.02)\times10^{42}
\left(L_{5100}/10^{44}\ {\rm erg\ s^{-1}}\right)^{1.157\pm0.005} \ {\rm erg\ s^{-1}}\)
from \citet{GreeneHo2005}, we obtain \(\log L_{5100}=42.657\pm0.007\) for the DESI epoch and \(\log L_{5100}=43.252\pm0.004\) for the MMT epoch, where \(L_{5100}\) denotes \(\lambda L_{\lambda}\) at rest-frame 5100~\AA. These formal uncertainties reflect the propagated measurement and fitting uncertainties, while the intrinsic scatter of the empirical relation is much larger, at approximately 0.2 dex, and therefore dominates the overall uncertainty in the inferred \(L_{5100}\). We estimated the bolometric luminosity as \(L_{\rm bol}=k_{5100}L_{5100}\), adopting \(k_{5100}=9\) as commonly used for type~1 AGNs \citep{Kaspi2000}. The Eddington luminosity was calculated as \(L_{\rm Edd}=1.26\times10^{38}(M_{\rm BH}/M_\odot)\ {\rm erg\ s^{-1}}\). Combining these luminosities with the respective single-epoch black hole mass estimates from Section~3.3, we infer Eddington ratios of $\lambda_{\rm Edd}\approx0.019$ for the DESI epoch and $\approx0.057$ for the MMT epoch. In other words, the source appears to have evolved from roughly 2\% to 6\% of the Eddington ratio between the two spectroscopic observations. This range falls within the few-percent regime commonly associated with changing-look or changing-state AGNs, where relatively modest changes in accretion power can lead to substantial variations in the ionizing continuum and the visibility of broad emission lines \citep[e.g.,][]{NodaDone2018,Guolo2021}. In a recent study of 20 BASS CLAGNs, \citet{Jana2025} found that all changing-look transitions were accompanied by changes in Eddington ratio and reported a median transition Eddington ratio of \(\sim 0.01\) for type~1 $\leftrightarrow$ 1.8/1.9/2 transitions. EP~J062132.0+623837 is therefore broadly consistent with the accretion-state changes observed in other CLAGNs, although its brighter-state Eddington ratio lies slightly above the median transition value reported by \citet{Jana2025}.

The inferred MIR reverberation scale is also broadly consistent with that expected from the source luminosity. Adopting the MMT-state luminosity, \(\log L_{5100}=43.252\), together with the bolometric correction \(L_{\rm bol}=9L_{5100}\) used above, gives \(\log(L_{\rm bol}/{\rm erg\,s^{-1}})\simeq44.21\). Using the empirical W1 and W2 lag--luminosity relations of \citet{Lyu2019}, this luminosity corresponds to expected observed-frame delays of approximately 90 and 116~d, respectively. These values are comparable to the characteristic optical--MIR delays inferred from both the DRW-based light-curve fitting and the independent ICCF analysis, supporting an interpretation in which the MIR variability traces dust reprocessing of changes in the accretion-powered continuum.

The X-ray spectrum of EP~J062132.0+623837 is well described by an AGN-like power-law continuum, but the measured photon index, \(\Gamma\sim1.5\)--1.6, is somewhat harder than the typical value of \(\Gamma\sim1.8\)--1.9 found for radio-quiet Seyfert galaxies \citep[e.g.,][]{NandraPounds1994,Lubinski2016}. This relatively hard spectrum may be related to the accretion state of
the source, although mild intrinsic absorption or additional spectral complexity cannot be ruled out with the present data. No compelling evidence for spectral-state changes is found between the different X-ray epochs. Using the best-fitting Swift/XRT photon index of $\Gamma=1.50$, we approximately convert the observed 0.3--10~keV luminosities to the rest-frame 2--10~keV band. Neglecting absorption in this band conversion, we obtain $L_{2-10}\simeq(2.3\text{--}5.9)\times10^{43}\ {\rm erg\ s^{-1}}$. For an illustrative bolometric correction of $k_{2-10}\equiv L_{\rm bol}/L_{2-10}=10\text{--}20$, the inferred bolometric luminosity is $L_{\rm bol}\simeq(2.3\text{--}11.7)\times10^{44}\ {\rm erg\ s^{-1}}$. For a characteristic black hole mass of $\log(M_{\rm BH}/M_\odot)\approx7.3$, this corresponds to $\lambda_{\rm Edd}\simeq0.09\text{--}0.47$. The bolometric correction varies substantially among AGNs \citep[e.g.,][]{Vasudevan2007,Vasudevan2009,Cheng2019}, so this X-ray-based estimate is approximate and is not used as a precise measure of the accretion state.

In summary, the strengthening of the broad Balmer emission lines and the long-term optical and MIR variability are most naturally interpreted as manifestations of changes in the accretion power. The contemporaneous X-ray detections establish that the nucleus was X-ray active during the EP discovery and follow-up interval, although the limited WXT baseline and the strongly varying exposure do not allow us to determine whether the X-ray emission underwent a corresponding long-term secular increase. In this context, EP~J062132.0+623837 fits naturally within the broader framework of changing-look or changing-state AGNs, while representing a relatively modest intermediate-type transition rather than an extreme turn-on event. As the optical continuum traces the accretion power, the optical light curves indicate that this source has undergone recurrent changes in appearance on timescales of years.

\subsection{Implications for Einstein Probe and X-ray-selected Changing-look AGNs}

EP~J062132.0+623837 illustrates an important aspect of the discovery space enabled by EP. The source is weak in individual WXT observations and was brought to attention through repeated wide-field monitoring and temporal binning. Importantly, the accumulated WXT exposure varies substantially with epoch because of the survey observing pattern. Consequently, the significance of an individual binned detection depends not only on the intrinsic source flux but also on the accumulated exposure. We therefore do not interpret the WXT detection history as evidence for a monotonic or systematic long-term X-ray brightening. Instead, the WXT measurements establish the presence of a weak and variable nuclear X-ray source during the available EP monitoring interval, while the subsequent FXT and \textit{Swift} observations establish that the source remained X-ray active during the follow-up period. This case also demonstrates that X-ray observations alone are insufficient to establish the physical nature of such a source. A robust classification was achieved only by combining the X-ray observations with long-term optical and MIR light curves and, most importantly, the pre- and post-discovery optical spectroscopy.


What makes EP~J062132.0+623837 especially notable is that it represents a CLAGN first identified through X-ray observations. Most CLAGNs discovered to date have been uncovered via optical variability, repeated optical spectroscopy, or MIR variability~\citep[e.g.,][]{MacLeod2016,Yang2018,Green2022,Sheng2020,Guo2024DESI}. By contrast, EP~J062132.0+623837 was initially identified as a candidate nuclear variable through its WXT detection and subsequently confirmed as a CLAGN through optical spectroscopic follow-up and examination of archival light curves.
The broader implication is that future EP nuclear transients will likely constitute a heterogeneous population. 
Some will be genuine new transients, whereas others will correspond to CLAGNs or related nuclear state changes identified through their X-ray variability. Disentangling these possibilities will require prompt localization and efficient multiwavelength follow-up, particularly optical spectroscopy and archival photometric context.

EP~J062132.0+623837 therefore provides a useful demonstration of how EP can expand the time-domain census of AGN activity. More specifically, it suggests that EP may provide an additional route to the discovery of CLAGNs from the X-ray band, complementing the more established optical and MIR selection routes. If similar objects are found in larger numbers, EP may help establish X-ray-selected CLAGNs as a promising sample in time-domain studies of black hole accretion.

\section{Conclusions}\label{sec:5}

We reported the EP discovery and multiwavelength follow-up of EP~J062132.0+623837, a nuclear X-ray source associated with the galaxy 2MASX J06213207+6238390 at \(z=0.084\). Our main conclusions are as follows.

\begin{enumerate}
\item EP~J062132.0+623837 was most clearly detected in 14-day binned WXT data in late 2025. Follow-up FXT observations localized the source to the nuclear region of the host galaxy, and the UKSSDC enhanced \textit{Swift}/XRT position is also consistent with the host-galaxy nucleus.

\item Long-term optical and MIR light curves from CRTS, ASAS-SN, ZTF, and \textit{WISE}/\textit{NEOWISE} show that the source had already undergone substantial multiwavelength variability over nearly two decades before the EP trigger. The EP event therefore occurred within a longer history of nuclear variability rather than as an isolated outburst.

\item Optical spectroscopic follow-up reveals a clear strengthening of the broad Balmer lines, marking a transition from a Type~1.8 state in the DESI epoch, with very weak broad H$\beta$, to a Type~1.5 state in the MMT epoch, where both broad H$\alpha$ and broad H$\beta$ are clearly visible. We therefore identify EP~J062132.0+623837 as a CLAGN first brought to attention by EP in the X-ray band.

\item The FXT and \textit{Swift}/XRT spectra are both well described by an absorbed redshifted power-law model, with photon indices of \(\Gamma \sim 1.5\)--1.6 and observed X-ray fluxes of order \(10^{-12}\ \mathrm{erg\ cm^{-2}\ s^{-1}}\). The source remained X-ray active throughout the initial post-discovery \textit{Swift} monitoring interval.

\item Using the broad H$\alpha$ luminosity and width, we estimated a characteristic black hole mass of \(\log(M_{\rm BH}/M_\odot)\approx 7.3\). The corresponding approximate Eddington ratios are \(\lambda_{\rm Edd}\approx 0.019\) and \(0.057\) for the DESI and MMT epochs, respectively, placing the source in the low-to-moderate accretion regime commonly found for changing-look or changing-state AGNs.
\end{enumerate}

EP~J062132.0+623837 illustrates that EP can identify variable nuclear X-ray sources whose physical nature can be revealed through multiwavelength follow-up. In this case, optical spectroscopy established the changing-look nature of the source, while the long-term optical and MIR variability provides additional context for its changing accretion state. More broadly, this case highlights the potential of EP to provide an additional route to the discovery of changing-look AGNs in the X-ray band, complementing optical and IR time-domain surveys.

\begin{acknowledgments}
We thank the anonymous referee for the valuable comments, which have significantly improved the manuscript.
This work is supported by the Strategic Priority Research Program of the Chinese Academy of Sciences (XDB0550200), the National Key Research and Development Program of China (2023YFA1608100), the National Natural Science Foundation of China (grants 12522303,12192221,12393814,12233008,12103048). This work is based on the data obtained with Einstein Probe (also known as "Tianguan" in Chinese), a space mission supported by the Strategic Priority Program on Space Science of Chinese Academy of Sciences, in collaboration with the European Space Agency, the Max-Planck-Institute for extraterrestrial Physics (Germany), and the Centre National d'\'Etudes Spatiales (CNES).

This work makes use of public data from the \textit{Neil Gehrels Swift Observatory}. We acknowledge the use of public data from the \textit{Swift} data archive provided by the High Energy Astrophysics Science Archive Research Center (HEASARC), which is operated by NASA's Goddard Space Flight Center.

This paper makes use of data products from the Dark Energy Spectroscopic Instrument (DESI). 

This work is based in part on observations obtained with the MMT Observatory.

This work makes use of data from the Zwicky Transient Facility (ZTF), the All-Sky Automated Survey for Supernovae (ASAS-SN), the Catalina Real-time Transient Survey (CRTS), and the \textit{Wide-field IR Survey Explorer} (\textit{WISE}) / \textit{NEOWISE}. \textit{WISE} is a joint project of the University of California, Los Angeles, and the Jet Propulsion Laboratory/California Institute of Technology, funded by the National Aeronautics and Space Administration. \textit{NEOWISE} is a project of the Jet Propulsion Laboratory/California Institute of Technology, funded by the Planetary Science Division of the National Aeronautics and Space Administration.

This research has made use of the NASA/IPAC Extragalactic Database (NED), SIMBAD, VizieR, and HEASARC.

\end{acknowledgments}

\software{
Astropy \citep[e.g.,][]{astropy2013,astropy2018,astropy2022},
HEASoft,
XSPEC \citep{Arnaud1996},
PypeIt \citep{Prochaska2020},
pPXF \citep{Cappellari2017},
NumPy \citep{numpy2020},
SciPy \citep{scipy2020},
Matplotlib \citep{Hunter2007},
PyCCF \citep{Sun2018PyCCF},
PyCALI \citep{Li2014},
custom Python scripts developed by the authors
}


\bibliography{sample701}{}
\bibliographystyle{aasjournalv7}



\clearpage
\onecolumngrid
\appendix
\section{Detailed Emission-line Measurements}
\label{app:linemeas}


Detailed emission-line measurements from the adopted DESI and MMT decompositions are listed in Table~\ref{tab:linemeas_full}. The fitting models and adopted line decompositions are described in Section~\ref{spec_analysis}. For the broad Balmer components, we report the integrated line fluxes and FWHM values used in the main text and in the subsequent black-hole mass estimates. For the narrow lines, the tabulated centroid, velocity offset, FWHM, and flux are measured from the summed narrow-line profiles of the adopted local decompositions.

Statistical uncertainties on the line fluxes and FWHM values were estimated from 500 Monte Carlo realizations. In each realization, Gaussian noise consistent with the spectral uncertainty array was added to the host-subtracted spectrum, and the same local decomposition was refitted. The quoted uncertainties correspond to the 16th--84th percentile ranges of the recovered parameter distributions.

\begin{center}
\parbox{0.95\textwidth}{
\centering
\refstepcounter{table}\label{tab:linemeas_full}
Table~\thetable. Emission-line measurements from the adopted local decompositions of the DESI and MMT spectra

\vspace{2mm}

\small
\setlength{\tabcolsep}{7pt}
\begin{tabular}{llccccc}
\hline
Epoch & Line & Component & Centroid & Velocity offset & FWHM & Flux \\
 &  &  & (\AA) & (km s$^{-1}$) & (km s$^{-1}$) & ($10^{-17}$ erg s$^{-1}$ cm$^{-2}$) \\
\hline
DESI & H$\alpha$ & Broad  & 6573.71 & 498.3 & $4786.0^{+80.8}_{-82.2}$ & $837.24^{+9.80}_{-10.09}$ \\
DESI & H$\alpha$ & Narrow & 6564.71 & 87.2  & $179.5^{+1.0}_{-0.8}$ & $2534.56^{+11.11}_{-11.27}$ \\
DESI & [N\,\textsc{ii}]$\lambda6548$ & Narrow & 6550.02 & 90.0  & $179.5^{+1.0}_{-0.9}$ & $415.47^{+3.86}_{-3.93}$ \\
DESI & [N\,\textsc{ii}]$\lambda6583$ & Narrow & 6585.34 & 85.9  & $179.5^{+0.9}_{-0.8}$ & $1254.27^{+5.90}_{-5.41}$ \\
DESI & [S\,\textsc{ii}]$\lambda6716$ & Narrow & 6718.43 & 88.7  & $179.5^{+1.0}_{-0.9}$ & $370.64^{+2.86}_{-2.84}$ \\
DESI & [S\,\textsc{ii}]$\lambda6731$ & Narrow & 6732.77 & 86.9  & $179.5^{+0.9}_{-0.9}$ & $388.36^{+3.83}_{-2.41}$ \\
DESI & H$\beta$ & Broad  & 4869.41 & 498.3 & $3068.0^{+345.9}_{-325.9}$ & $98.81^{+7.04}_{-5.81}$ \\
DESI & H$\beta$ & Narrow & 4862.25 & 56.6  & $179.5^{+2.9}_{-2.5}$ & $558.17^{+5.23}_{-5.32}$ \\
DESI & [O\,\textsc{iii}]$\lambda4959$ & Narrow & 4958.94 & 1.6   & $323.0^{+3.6}_{-3.5}$ & $675.03^{+2.25}_{-2.40}$ \\
DESI & [O\,\textsc{iii}]$\lambda5007$ & Narrow & 5006.87 & 1.6   & $323.0^{+3.6}_{-3.8}$ & $2031.03^{+6.76}_{-7.22}$ \\
\hline
MMT & H$\alpha$ & Broad  & 6570.82 & 366.3 & $3524.0^{+14.8}_{-12.7}$ & $4090.46^{+12.67}_{-14.82}$ \\
MMT & H$\alpha$ & Narrow & 6566.95 & 189.6 & $180.7^{+0.8}_{-0.9}$ & $1866.16^{+6.94}_{-6.86}$ \\
MMT & [N\,\textsc{ii}]$\lambda6548$ & Narrow & 6552.19 & 189.6 & $240.9^{+1.6}_{-1.6}$ & $310.34^{+3.20}_{-3.00}$ \\
MMT & [N\,\textsc{ii}]$\lambda6583$ & Narrow & 6587.73 & 194.9 & $240.9^{+1.6}_{-1.2}$ & $946.71^{+6.69}_{-4.39}$ \\
MMT & [S\,\textsc{ii}]$\lambda6716$ & Narrow & 6721.62 & 231.0 & $180.7^{+2.6}_{-2.4}$ & $302.93^{+1.99}_{-2.04}$ \\
MMT & [S\,\textsc{ii}]$\lambda6731$ & Narrow & 6735.92 & 227.1 & $241.0^{+3.4}_{-3.7}$ & $305.79^{+2.24}_{-1.91}$ \\
MMT & H$\beta$ & Broad  & 4867.27 & 366.3 & $4124.0^{+55.2}_{-58.5}$ & $872.60^{+9.00}_{-7.60}$ \\
MMT & H$\beta$ & Narrow & 4866.43 & 314.2 & $241.0^{+2.3}_{-2.3}$ & $550.73^{+4.24}_{-4.13}$ \\
MMT & [O\,\textsc{iii}]$\lambda4959$ & Narrow & 4963.63 & 285.1 & $361.5^{+2.8}_{-2.3}$ & $676.19^{+2.11}_{-2.25}$ \\
MMT & [O\,\textsc{iii}]$\lambda5007$ & Narrow & 5011.60 & 285.1 & $361.5^{+2.7}_{-2.3}$ & $2034.53^{+6.34}_{-6.76}$ \\
\hline
\end{tabular}

\vspace{2mm}

\begin{minipage}{0.92\textwidth}
\footnotesize
\textit{Notes.}
Broad-component fluxes and FWHM values are those adopted in the main text, while narrow-component measurements are obtained from the summed narrow-line profiles of the adopted local decompositions. Quoted uncertainties on the line fluxes and FWHM values correspond to the 16th--84th percentile ranges of the Monte Carlo distributions described above.
Centroids are given in the rest frame, velocity offsets are measured relative to the corresponding laboratory wavelengths, FWHM values are in km s$^{-1}$, and fluxes are in units of $10^{-17}$ erg s$^{-1}$ cm$^{-2}$.
Formal uncertainties are not reported for the centroids and velocity offsets because these quantities are auxiliary and more model-dependent than the integrated line fluxes and widths.
For broad H$\beta$, the centroid and velocity offset are tied to those of the better-constrained broad H$\alpha$ component and should therefore not be interpreted as independent measurements.
\end{minipage}
}
\end{center}

\end{document}